\UseRawInputEncoding
\documentclass{ws-ijgmmp}

\begin{document}

\markboth{Zi-Hua Weng}
{Two incompatible types of invariants in the octonion spaces}

%
\catchline{}{}{}{}{}
%

\title{Two incompatible types of invariants in the octonion spaces
}

\author{Zi-Hua Weng
}
\address{School of Aerospace Engineering, Xiamen University, Xiamen, China
\\
College of Physical Science and Technology, Xiamen University, Xiamen, China
\\
\email{xmuwzh@xmu.edu.cn
}
}


\maketitle

\begin{history}
\received{(Day Month Year)}
\revised{(Day Month Year)}
\end{history}

\begin{abstract}
  The paper aims to study some invariants and conservation laws relevant to electromagnetic and gravitational fields, by means of the rotational transformations of octonion coordinate systems. The scholars utilize the octonions to analyze the electromagnetic and gravitational fields simultaneously, including the octonion field potential, field strength, field source, linear momentum, angular momentum, torque and force. When the octonion coordinate system transforms rotationally, the vector part of one octonion may alter, while the scalar part of the octonion will remain unchanged. This property allows one to deduce a few invariants, such as the scalar part of octonion radius vector, speed of light, and norm of octonion radius vector. These invariants are the basic postulates for the Galilean transformation and Lorentz transformation. Further, from the rotation transform of octonion coordinate systems, it is capable of deducing several invariants, including the mass, energy and power relevant to gravitational fields, in one octonion space $\mathbb{O}$. And the term relevant to the electric charge will transform with the rotation of coordinate systems. In another octonion space $\mathbb{O}_u$, it is capable of inferring a few invariants, including the electric charge related to electromagnetic fields. And the terms relevant to the mass and energy will vary with the rotation of octonion coordinate systems. So the invariants are divided into two different groups. In particular, the mass conservation law and energy conservation law can be effective simultaneously. But the charge conservation law and mass conservation law are unable to be valid simultaneously, in the strict sense. It is beneficial to further understand the laws of conservation.
\end{abstract}


\section{\label{sec:level1}Introduction}

Is there a new law of conservation? Can the material medium contribute to the laws of conservation? Can the mass conservation law and charge conservation law be effective simultaneously? For a long time, these difficult problems have puzzled and attracted scholars. It was not until the emergence of octonion field theory (short for the field theory described with the octonions) that these questions were answered in part. When the octonion coordinate system transforms rotationally, the vector part of one octonion will alter, while the scalar part of one octonion will remain unchanged. Starting from the characteristics of rotation transform of octonion coordinate systems, it is able to deduce several invariants relevant to the electromagnetic and gravitational fields, such as the speed of light, mass, energy, and electric-charge. Strictly speaking, the mass conservation law and energy conservation law can be effective simultaneously. However, the charge conservation law and mass conservation law cannot be valid simultaneously. This expands the understanding of the properties of invariants and conservation laws.

The existing field theories have some defects in the exploration of several invariants and others relevant to the physical quantities. Through the analysis, it is found that there are a few shortcomings in the study of conservation laws relevant to the physical quantities.

1) Single set of invariants. The classical field theory holds that all of conservation laws belong to the same group, in the electromagnetic and gravitational fields. And all of conservation laws are effective simultaneously. In particular, the mass conservation law, energy conservation law, and charge conservation law and others must be established simultaneously. However, this is a transcendental hypothesis, which has not yet been validated by experiments so far.

2) Independent of material media. According to the classical field theory, the material media have no effect on the invariants of physical quantities. The material medium has no contribution to the laws of conservation, in the electromagnetic and gravitational fields. In particular, the energy conservation law, gauge equations, and power conservation law (in Section 5) in material media are the same as they are in vacuum, respectively. This is an assumed guess.

3) Experimental limitations. Based on the experiments, some scholars put forward the mass conservation law and the charge conservation law. Subsequently, a few scholars come up with the energy conservation law. Overemphasizing the importance of experiments limits the discovery of new laws of conservation. The classical field theory is unable to summarize new conservation laws from the existing experiments. At present, some regular physical phenomena are considered to be random and disordered.

It is in sharp contrast with the above, the application of octonions is capable of solving several problems of invariants in the gravitational and electromagnetic theories, settling a few puzzles derived from the existing theories. Making use of the octonion field theory, it is able to deeply explore some characteristics of invariants and conservation laws related to physical quantities.

J. C. Maxwell first utilized the quaternions and vector analysis to describe the physical properties of electromagnetic fields. Nowadays, some scholars \cite{mironov} apply the quaternions \cite{morita} and octonions \cite{deleo} to research the electromagnetic fields \cite{tanisli1,demir}, gravitational fields \cite{rawat}, relativity \cite{moffat}, quantum mechanics \cite{gogberashvili,bernevig}, dark matters \cite{furui}, astrophysical jets, strong nuclear fields \cite{furey,chanyal2}, weak nuclear fields \cite{majid,farrill}, black holes \cite{bossard}, invariants, equilibrium equations, continuity equations, hydromechanics \cite{tanisli,chanyal3},
optics \cite{liu}, and computer graphics \cite{goldman} and others.

In the gravitational and electromagnetic theories described with the algebra of octonions, the exploration of invariants and conservation laws and others possesses a few advantages as follows.

1) Two group of invariants. In the gravitational and electromagnetic fields, the conservation laws are divided into two different types of groups, according to the octonion field theory. The conservation laws in the same group can be effective simultaneously. But the laws of conservation among different types of groups cannot be valid simultaneously. In particular, the mass conservation law and energy conservation law belong to the same group, so both of them can be effective simultaneously. But the mass conservation law and charge conservation law belong to two different types of groups, and they can't be effective simultaneously.

2) Contribution of material media. The material medium does not contribute to several invariants, such as, the octonion radius vector, velocity, and field potential. However, the material media are able to exert an influence on some invariants, in the octonion field strength, field source, linear momentum, angular momentum, torque, and force. The material medium has a certain contribution to some conservation laws. In particular, the mass conservation law, energy conservation law, and charge conservation law and others within the material medium are different from those in vacuum, respectively.

3) Theoretical predictions. The rotational transformations of octonion coordinate systems are able to deduce some invariants, including the norm of octonion radius vector, speed of light, and scalar part of octonion radius vector. These invariants are the basic postulates for the Galilean transformation and Lorentz transformation. In terms of the physical quantities in the gravitational fields, it is able to infer the mass conservation law, energy conservation law, and power conservation law and others. For the physical quantities in the electromagnetic fields, it is capable of deducing the charge conservation law and others. The octonion field theory can put forward some new laws of conservation.

When the rotational transformation occurs in the octonion coordinate systems, the vector part of the octonion physical quantity will change, while its scalar part remains unchanged. This property is able to deduce some invariants relevant to gravitational fields, including the scalar part of octonion radius vector, speed of light, norm of octonion radius vector, mass, energy, and power and so on, in the octonion space $\mathbb{O}$ . Similarly, it is capable of inferring a few invariants related to electromagnetic fields, including the electric-charge and so forth, in the octonion space $\mathbb{O}_u$ . Because the vectors in the two octonion spaces, $\mathbb{O}$ and $\mathbb{O}_u$ , are different from each other, the invariants are divided into two different types of groups, in the gravitational and electromagnetic fields. In particular, the mass conservation law and energy conservation law can be valid simultaneously. But the charge conservation law and mass conservation law cannot be effective simultaneously, in the strict sense. It is beneficial to deepen the further understanding of the properties of invariants and laws of conservation.

\section{\label{sec:level1}Octonion spaces}

The algebra of octonions can be applied to describe the physical properties of electromagnetic and gravitational fields simultaneously. The octonion space is decomposed into some subspaces independent of each other, including $\mathbb{H}_g$ and $\mathbb{H}_{em}$ . The subspace $\mathbb{H}_g$ is one quaternion space, which is able to explore the physical properties of gravitational fields. And the second subspace $\mathbb{H}_{em}$ is one four-dimension space, which is capable of studying the physical properties of electromagnetic fields.

In the quaternion space, $\mathbb{H}_g$ , for gravitational fields, the basis vector is $\textbf{i}_j$ , and the coordinate value is $r_j$ . The radius vector is, $\mathbb{R}_g = i \textbf{i}_0 r_0 + \Sigma \textbf{i}_k r_k$ . The velocity is, $\mathbb{V}_g = i \textbf{i}_0 v_0 + \Sigma \textbf{i}_k v_k$ . The gravitational potential is, $\mathbb{A}_g = i \textbf{i}_0 a_0 + \Sigma \textbf{i}_k a_k$. The gravitational strength is, $\mathbb{F}_g = \textbf{i}_0 f_0 + \Sigma \textbf{i}_k f_k$ . The gravitational source is, $\mathbb{S}_g = i \textbf{i}_0 s_0 + \Sigma \textbf{i}_k s_k$. Herein $r_0 = v_0 t$. $v_0$ is the speed of light, $t$ is the time. $r_j , v_j , a_j , f_0$, and $s_j$ are all real. $f_k$ is complex. $i$ is the imaginary unit. $j = 0, 1, 2, 3$. $k = 1, 2, 3$. $\textbf{i}_0 = 1$.

In the second subspace, $\mathbb{H}_{em}$ , for electromagnetic fields, the basis vector is $\textbf{I}_j$ , and the coordinate value is $R_j$ . The radius vector is, $\mathbb{R}_e = i \textbf{I}_0 R_0 + \Sigma \textbf{I}_k R_k$ . The velocity is, $\mathbb{V}_e = i \textbf{I}_0 V_0 + \Sigma \textbf{I}_k V_k$ . The electromagnetic potential is, $\mathbb{A}_e = i \textbf{I}_0 A_0 + \Sigma \textbf{I}_k A_k$ . The electromagnetic strength is, $\mathbb{F}_e = \textbf{I}_0 F_0 + \Sigma \textbf{I}_k F_k$ . The electromagnetic source is, $\mathbb{S}_e = i \textbf{I}_0 S_0 + \Sigma \textbf{I}_k S_k$ . Herein $R_j$ , $V_j$ , $A_j$ , $F_0$ , and $S_j$ are all real. $F_k$ is complex. $\textbf{I}_k = \textbf{i}_k \circ \textbf{I}_0$ . $\circ$ indicates the octonion multiplication.

As a result, in the octonion space, $\mathbb{O}$ , for the gravitational and electromagnetic fields, the octonion radius vector is written as, $\mathbb{R} = \mathbb{R}_g + k_{eg} \mathbb{R}_e$ , and the octonion velocity is written as, $\mathbb{V} = \mathbb{V}_g + k_{eg} \mathbb{V}_e$ . The octonion field potential can be written as, $\mathbb{A} = \mathbb{A}_g + k_{eg} \mathbb{A}_e$ , and the octonion field strength can be written as, $\mathbb{F} = \mathbb{F}_g + k_{eg} \mathbb{F}_e$. Herein $\mathbb{V} = \partial \mathbb{R} / \partial t$ . $\mathbb{F} = \lozenge \circ \mathbb{A}$ , with $\lozenge = \Sigma \textbf{i}_j \partial / \partial r_j $ . $\nabla = \Sigma \textbf{i}_k \partial / \partial r_k $ . $k_{eg}$ is the coefficient, to meet the demand for dimensional homogeneity.

In the rotation transform of octonion coordinate systems, the scalar parts of octonions are the invariants. These invariants are the basic postulates for the Galilean transformation and Lorentz transformation and so forth.

\section{\label{sec:level1}Coordinate transformations}

In the octonion space $\mathbb{O}$ , an octonion physical quantity possesses eight independent components, including one scalar component and seven vectorial components. When an octonion physical quantity is transformed rotationally from one octonion coordinate system $\zeta$ to another octonion coordinate system $\eta$ , the vectorial components of this physical quantity will vary, while the scalar component of the physical quantity remains unchanged. Starting from the characteristics of rotation transform of octonion coordinate systems, it is able to draw out several invariants, including the scalar part of octonion radius vector, scalar part of octonion velocity, and norm of octonion radius vector and so forth.

In a vacuum without considering the contribution of any material medium, when an octonion physical quantity is transformed rotationally from one octonion coordinate system, $\alpha$ , to another octonion coordinate system, $\beta$ , the octonion radius vector, $\mathbb{R}$ , and octonion velocity, $\mathbb{V}$ , in the octonion coordinate system, $\alpha$, can transform into the octonion radius vector, $\mathbb{R}^\prime$ , and octonion velocity, $\mathbb{V}^\prime$ , in the octonion coordinate system, $\beta$ , respectively.

In the octonion coordinate system, $\alpha$ , the octonion radius vector, $\mathbb{R}$ , and octonion velocity, $\mathbb{V}$ , can be written as follows,
\begin{eqnarray}
&& \mathbb{R} = i \textbf{i}_0 r_0 + \Sigma \textbf{i}_k r_k + k_{eg} ( i \textbf{I}_0 R_0 + \Sigma \textbf{I}_k R_k ) ~ ,
\\
&& \mathbb{V} = i \textbf{i}_0 v_0 + \Sigma \textbf{i}_k v_k + k_{eg} ( i \textbf{I}_0 V_0 + \Sigma \textbf{I}_k V_k ) ~ .
\end{eqnarray}

In the octonion coordinate system, $\beta$ , the octonion radius vector, $\mathbb{R}^\prime$ , and octonion velocity, $\mathbb{V}^\prime$ , can be written as follows,
\begin{eqnarray}
&& \mathbb{R}^\prime = i \textbf{i}_0^\prime r_0^\prime + \Sigma \textbf{i}_k^\prime r_k^\prime + k_{eg} ( i \textbf{I}_0^\prime R_0^\prime + \Sigma \textbf{I}_k^\prime R_k^\prime ) ~ ,
\\
&& \mathbb{V}^\prime = i \textbf{i}_0^\prime v_0^\prime + \Sigma \textbf{i}_k^\prime v_k^\prime + k_{eg} ( i \textbf{I}_0^\prime V_0^\prime + \Sigma \textbf{I}_k^\prime V_k^\prime ) ~ ,
\end{eqnarray}
where $\textbf{i}_0^\prime = 1$.

1) Invariants. When the octonion coordinate system $\alpha$ is transformed into another octonion coordinate system $\beta$, the scalar parts of octonion radius vector $\mathbb{R}$ and octonion velocity $\mathbb{V}$ both are invariants under the coordinate transformation. From Eqs.(1)-(4), it is able to achieve two invariants as follows,
\begin{eqnarray}
&& r_0^\prime = r_0 ~ ,
\\
&& v_0^\prime = v_0 ~ .
\end{eqnarray}

Similarly, the norm of octonion radius vector $\mathbb{R}$ is one scalar, which is also an invariance under the coordinate transformation. Consequently, From Eqs.(1) and (3), there is,
\begin{eqnarray}
&&  ( \mathbb{R}^\prime )^\ast \circ \mathbb{R}^\prime = \mathbb{R}^\ast \circ \mathbb{R} ~,
\end{eqnarray}
where $\ast$ denotes the conjugate of octonion. The constant $k_{eg}$ is also an invariance under the coordinate transformation.

In a vacuum without considering the contribution of any material medium, these invariants can be used as the basic postulates for the Galilean transformation and Lorentz transformation and so forth.

2) Galilean transformation. In the octonion space $\mathbb{O}$ , from Eqs.(5) and (6), it is able to achieve two basic postulates of Galilean transformation as follows,
\begin{eqnarray}
&& r_0^\prime = r_0 ~ ,~~ v_0^\prime = v_0 ~ .
\end{eqnarray}

From the two basic postulates, it is capable of deducing the Galilean transformation with eight coordinate values. In case the octonion space $\mathbb{O}$ is degenerated into the quaternion space $\mathbb{H}_g$ , the Galilean transformation with eight coordinate values \cite{weng1} will be reduced into the Galilean transformation in the classical field theory.

3) Lorentz transformation. In the octonion space $\mathbb{O}$ , from Eqs.(6) and (7), it is able to achieve two basic postulates of Lorentz transformation as follows,
\begin{eqnarray}
&&  ( \mathbb{R}^\prime )^\ast \circ \mathbb{R}^\prime = \mathbb{R}^\ast \circ \mathbb{R} ~,~~ v_0^\prime = v_0 ~ .
\end{eqnarray}

From the two basic postulates, it is capable of deducing the Lorentz transformation with eight coordinate values (see Ref.\cite{weng1}). When the scalar part of octonion radius vector plays a major role, two basic postulates, Eq.(9), derived from the octonion space $\mathbb{O}$ , will be approximately degenerate into two basic postulates, Eq.(8). When the octonion space $\mathbb{O}$ is simplified into the quaternion space $\mathbb{H}_g$ , two basic postulates, Eq.(9), in the octonion space $\mathbb{O}$ , will be approximately degenerate into two basic postulates in the special theory of relativity.

This method can be extended to some invariants in the octonion field equations, including the mass, energy, and electric-charge and others. Further, it is found that the mass conservation law and charge conservation law are unable to be effective simultaneously.

\section{\label{sec:level1}Field equations in vacuum}

In the case of without considering the contribution of any material medium, it is able to achieve the octonion field equations relevant to the gravitational and electromagnetic fields \cite{weng2}.

The octonion field source, $\mu \mathbb{S} = - ( i \mathbb{F} / v_0 + \lozenge )^\ast \circ \mathbb{F}$ , can be rewritten as, $\mu \mathbb{S} = \mu_g \mathbb{S}_g + k_{eg} \mu_e \mathbb{S}_e - i \mathbb{F}^\ast \circ \mathbb{F} / v_0 $ . According to the coefficient $k_{eg}$ and basis vector, there are, $\mu_g \mathbb{S}_g = - \lozenge^\ast \circ \mathbb{F}_g $ , and $\mu_e \mathbb{S}_e = - \lozenge^\ast \circ \mathbb{F}_e $ . The former is the gravitational equations in the quaternion space $\mathbb{H}_g$ , while the latter is the electromagnetic equations in the second subspace $\mathbb{H}_{em}$ . Herein $\mu$ is one coefficient. $\mu_g$ is the gravitational constant. $\mu_e$ is the electromagnetic constant. $k_{eg}^2 = \mu_g / \mu_e$ .

The octonion linear momentum, $\mathbb{P} = \mu \mathbb{S} / \mu_g$ , is written as, $\mathbb{P} = \mathbb{P}_g + k_{eg} \mathbb{P}_e$ . Herein $\mathbb{P}_g$ and $\mathbb{P}_e$ are situated in two subspaces, $\mathbb{H}_g$ and $\mathbb{H}_{em}$ , respectively. For one single particle, $s_0 = m_g v_0$ , with $m_g$ being the gravitational mass. $S_0 = q V_0$ . $q$ is the electric charge. $V_0$ is called as the second speed of light, temporarily.

Further the octonion angular momentum, $\mathbb{L}$ , can be written as,
\begin{eqnarray}
&&  \mathbb{L} = ( \mathbb{R} + k_{rx} \mathbb{X} )^\times \circ \mathbb{P} ~ ,
\end{eqnarray}
where $\times$ indicates the complex conjugate. $k_{rx}$ is one coefficient to satisfy the requirement of dimensional homogeneity, with $k_{rx} = 1 / v_0$ . $\mathbb{X}$ is the octonion integrating function of field potential. By comparison with the octonion radius vector $\mathbb{R}$ , the term $k_{rx} \mathbb{X}$ can be neglected in general. The octonion angular momentum can be rewritten as, $\mathbb{L} = \mathbb{L}_g + k_{eg} \mathbb{L}_e$. $\mathbb{L}_g$ and $\mathbb{L}_e$ are situated in the subspaces, $\mathbb{H}_g$ and $\mathbb{H}_{em}$ , respectively. $\mathbb{L}_g = L_{10} + i \textbf{L}_1^i + \textbf{L}_1$ . $\mathbb{L}_e = \textbf{L}_{20} + i \textbf{L}_2^i + \textbf{L}_2$ . $\textbf{L}_1$ is the angular momentum. $\textbf{L}_1^i$ is called as the mass moment, temporarily. $\textbf{L}_2^i$ is the electric moment, and $\textbf{L}_2$ is the magnetic moment. $\textbf{L}_1 = \Sigma L_{1k} \textbf{i}_k$ , $\textbf{L}_1^i = \Sigma L_{1k}^i \textbf{i}_k$ . $\textbf{L}_2 = \Sigma L_{2k} \textbf{I}_k$, $\textbf{L}_2^i = \Sigma L_{2k}^i \textbf{I}_k$ . $\textbf{L}_{20} = \textbf{I}_0 L_{20}$. $L_{1j} , L_{2j} , L_{1k}^i$ , and $L_{2k}^i$ are all real.

The octonion torque, $\mathbb{W}$ , is written as,
\begin{eqnarray}
&&  \mathbb{W} = - v_0 ( i \mathbb{F} / v_0 + \lozenge ) \circ \{ ( i \mathbb{V}^\times / v_0 ) \circ \mathbb{L} \} ~ ,
\end{eqnarray}
where the octonion torque can be rewritten as, $\mathbb{W} = \mathbb{W}_g + k_{eg} \mathbb{W}_e$ . $\mathbb{W}_g$ and $\mathbb{W}_e$ are situated in the subspaces, $\mathbb{H}_g$ and $\mathbb{H}_{em}$ , respectively. $\mathbb{W}_g = i W_{10}^i + W_{10} + i \textbf{W}_1^i + \textbf{W}_1$. $\mathbb{W}_e = i \textbf{W}_{20}^i + \textbf{W}_{20} + i \textbf{W}_2^i + \textbf{W}_2$ . $W_{10}^i$ is the energy. $\textbf{W}_1^i$ is the torque, including the gyroscopic torque, $\nabla ( \textbf{v} \cdot \textbf{L}_1 )$. $\textbf{W}_{20}^i$ is the second energy. $\textbf{W}_2^i$ is the second torque. $\textbf{W}_1 = \Sigma W_{1k} \textbf{i}_k$ , $\textbf{W}_1^i = \Sigma W_{1k}^i \textbf{i}_k$ . $\textbf{W}_2 = \Sigma W_{2k} \textbf{I}_k$ , $\textbf{W}_2^i = \Sigma W_{2k}^i \textbf{I}_k$ , $\textbf{W}_{20}^i = W_{20}^i \textbf{I}_0$. $\textbf{W}_{20} = W_{20} \textbf{I}_0$ . $\textbf{v} = \Sigma v_k \textbf{i}_k$ . $W_{1j} , W_{2j} , W_{1j}^i$ , and $W_{2j}^i$ are all real.

The octonion force, $\mathbb{N}$ , is written as,
\begin{eqnarray}
&&  \mathbb{N} = - ( i \mathbb{F} / v_0 + \lozenge ) \circ \{ ( i \mathbb{V}^\times / v_0 ) \circ \mathbb{W} \} ~ ,
\end{eqnarray}
where the octonion force can be rewritten as, $\mathbb{N} = \mathbb{N}_g + k_{eg} \mathbb{N}_e$ . $\mathbb{N}_g$ and $\mathbb{N}_e$ are situated in the subspaces, $\mathbb{H}_g$ and $\mathbb{H}_{em}$ , respectively. $\mathbb{N}_g = i N_{10}^i + N_{10} + i \textbf{N}_1^i + \textbf{N}_1$. $\mathbb{N}_e = i \textbf{N}_{20}^i + \textbf{N}_{20} + i \textbf{N}_2^i + \textbf{N}_2$ . $N_{10}$ is the power. $\textbf{N}_1^i$ is the force, including the Magnus force, $ \nabla ( \partial L_{10} / \partial t)$. $\textbf{N}_{20}$ is the second power. $\textbf{N}_2^i$ is the second force. $\textbf{N}_1 = \Sigma N_{1k} \textbf{i}_k$, $\textbf{N}_1^i = \Sigma N_{1k}^i \textbf{i}_k$ . $\textbf{N}_2 = \Sigma N_{2k} \textbf{I}_k$ , $\textbf{N}_2^i = \Sigma N_{2k}^i \textbf{I}_k$ , $\textbf{N}_{20}^i = N_{20}^i \textbf{I}_0$ . $\textbf{N}_{20} = N_{20} \textbf{I}_0$ . $N_{1j} , N_{2j} , N_{1j}^i$ , and $N_{2j}^i$ are all real.

\section{\label{sec:level1}Gravitational fields in vacuum}

In a vacuum without considering the contribution of any material medium, when an octonion coordinate system, $\alpha$ , is transformed rotationally to another octonion coordinate system, $\beta$, the physical quantities, $\mathbb{A}$ , $\mathbb{F}$ , $\mathbb{P}$ , $\mathbb{L}$ , $\mathbb{W}$ , and $\mathbb{N}$ , can transform into the physical quantities, $\mathbb{A}^\prime$ , $\mathbb{F}^\prime$ , $\mathbb{P}^\prime$ , $\mathbb{L}^\prime$ , $\mathbb{W}^\prime$ , and $\mathbb{N}^\prime$ , respectively.

In the octonion coordinate system $\alpha$ , the octonion physical quantities, $\mathbb{A}$ , $\mathbb{F}$ , $\mathbb{P}$ , $\mathbb{L}$ , $\mathbb{W}$, and $\mathbb{N}$ , can be written as follows,
\begin{eqnarray}
&& \mathbb{A} = i a_0 + \Sigma \textbf{i}_k a_k + k_{eg} ( i \textbf{I}_0 A_0 + \Sigma \textbf{I}_k A_k ) ~ ,
\\
&& \mathbb{F} = f_0 + \Sigma \textbf{i}_k f_k + k_{eg} ( \textbf{I}_0 F_0 + \Sigma \textbf{I}_k F_k ) ~ ,
\\
&& \mathbb{P} = i s_0 + \Sigma \textbf{i}_k s_k + k_{eg}^{-1} ( i \textbf{I}_0 S_0 + \Sigma \textbf{I}_k S_k ) ~ ,
\\
&& \mathbb{L} = L_{10} + i \Sigma \textbf{i}_k L_{1k}^i + \Sigma \textbf{i}_k L_{1k}
\nonumber
\\
&&~~~~~~~
+ k_{eg} ( \textbf{I}_0 L_{20} + i \Sigma \textbf{I}_k L_{2k}^i + \Sigma \textbf{I}_k L_{2k}  ) ~ ,
\\
&& \mathbb{W} = i W_{10}^i + W_{10} + i \Sigma \textbf{i}_k W_{1k}^i + \Sigma \textbf{i}_k W_{1k}
\nonumber
\\
&&~~~~~~~
+ k_{eg} ( i \textbf{I}_0 W_{20}^i + \textbf{I}_0 W_{20} + i \Sigma \textbf{I}_k W_{2k}^i + \Sigma \textbf{I}_k W_{2k} ) ~ ,
\\
&& \mathbb{N} = i N_{10}^i + N_{10} + i \Sigma \textbf{i}_k N_{1k}^i + \Sigma \textbf{i}_k N_{1k}
\nonumber
\\
&&~~~~~~~
+ k_{eg} ( i \textbf{I}_0 N_{20}^i + \textbf{I}_0 N_{20} + i \Sigma \textbf{I}_k N_{2k}^i + \Sigma \textbf{I}_k N_{2k} ) ~ .
\end{eqnarray}

In the octonion coordinate system $\beta$ , the octonion physical quantities, $\mathbb{A}^\prime$ , $\mathbb{F}^\prime$, $\mathbb{P}^\prime$ , $\mathbb{L}^\prime$ , $\mathbb{W}^\prime$ , and $\mathbb{N}^\prime$ , can be written as, respectively,
\begin{eqnarray}
&& \mathbb{A}^\prime = i a_0^\prime + \Sigma \textbf{i}_k^\prime a_k^\prime + k_{eg} ( i \textbf{I}_0^\prime A_0^\prime + \Sigma \textbf{I}_k^\prime A_k^\prime ) ~ ,
\\
&& \mathbb{F}^\prime = f_0^\prime + \Sigma \textbf{i}_k^\prime f_k^\prime + k_{eg} ( \textbf{I}_0^\prime F_0^\prime + \Sigma \textbf{I}_k^\prime F_k^\prime ) ~ ,
\\
&& \mathbb{P}^\prime = i s_0^\prime + \Sigma \textbf{i}_k^\prime s_k^\prime + k_{eg}^{-1} ( i \textbf{I}_0^\prime S_0^\prime + \Sigma \textbf{I}_k^\prime S_k^\prime ) ~ ,
\\
&& \mathbb{L}^\prime = L_{10}^\prime + i \Sigma \textbf{i}_k^\prime L_{1k}^{i \prime} + \Sigma \textbf{i}_k^\prime L_{1k}^\prime
\nonumber
\\
&&~~~~~~~
+ k_{eg} ( \textbf{I}_0^\prime L_{20}^\prime + i \Sigma \textbf{I}_k^\prime L_{2k}^{i \prime} + \Sigma \textbf{I}_k^\prime L_{2k}^\prime  ) ~ ,
\\
&& \mathbb{W}^\prime = i W_{10}^{i \prime} + W_{10}^\prime + i \Sigma \textbf{i}_k^\prime W_{1k}^{i \prime} + \Sigma \textbf{i}_k^\prime W_{1k}^\prime
\nonumber
\\
&&~~~~~~~
+ k_{eg} ( i \textbf{I}_0^\prime W_{20}^{i \prime} + \textbf{I}_0^\prime W_{20}^\prime + i \Sigma \textbf{I}_k^\prime W_{2k}^{i \prime} + \Sigma \textbf{I}_k^\prime W_{2k}^\prime ) ~ ,
\\
&& \mathbb{N}^\prime = i N_{10}^{i \prime} + N_{10}^\prime + i \Sigma \textbf{i}_k^\prime N_{1k}^{i \prime} + \Sigma \textbf{i}_k^\prime N_{1k}^\prime
\nonumber
\\
&&~~~~~~~
+ k_{eg} ( i \textbf{I}_0^\prime N_{20}^{i \prime} + \textbf{I}_0^\prime N_{20}^\prime + i \Sigma \textbf{I}_k^\prime N_{2k}^{i \prime} + \Sigma \textbf{I}_k^\prime N_{2k}^\prime ) ~ .
\end{eqnarray}

When the octonion physical quantities are transformed rotationally from one octonion coordinate system, $\alpha$ , to another octonion coordinate system, $\beta$ , each of scalar parts of some octonion physical quantities, relevant to gravitational fields, is invariant under the coordinate transformations. From Eqs.(13)-(24), there are eight invariants as follows,
\begin{eqnarray}
&& a_0^\prime = a_0 ~ ,
\\
&& f_0^\prime = f_0 ~ ,
\\
&& s_0^\prime = s_0 ~ ,
\\
&& L_{10}^\prime = L_{10} ~ ,
\\
&& W_{10}^{i \prime} = W_{10}^i ~ ,
\\
&& W_{10}^\prime = W_{10} ~ ,
\\
&& N_{10}^{i \prime} = N_{10}^i ~ ,
\\
&& N_{10}^\prime = N_{10} ~ .
\end{eqnarray}

From Eqs.(6) and (27), it is able to achieve the mass conservation law. From Eq.(29), it will deduce the energy conservation law. From Eq.(32), it can infer the power conservation law. Eqs.(25), (26), and (28) imply that the scalar part of gravitational potential, the scalar part of gravitational strength, and the dot product of angular momenta are the invariants independent of each other. Eqs.(30) and (31) state that the divergence of angular momenta and the divergence of torques are invariants also.

The above analysis shows that there are multiple conservation laws, in the octonion space $\mathbb{O}$ . And that these conservation laws are able to be established simultaneously. For instance, the mass conservation law, energy conservation law, and power conservation law can be effective simultaneously (Table 1).

The above method can be extended from the gravitational fields in vacuum to the electromagnetic fields in vacuum.

\begin{table}[h]
\centering
\caption{Some invariants and conservation laws relevant to gravitational fields, in the octonion space $\mathbb{O}$ without considering the contribution of any material medium.}
\label{tab:2}       
\begin{tabular}{@{}lll@{}}
\hline\noalign{\smallskip}
physical quantity               &   invariants                              &   conservation laws                              \\
\noalign{\smallskip}\hline\noalign{\smallskip}
$\mathbb{R}$                    &   $r_0^\prime = r_0$                      &   conservation law of the scalar part of         \\
                                &                                           &   ~~~~~octonion radius vector                    \\
$\mathbb{V}$                    &   $v_0^\prime = v_0$                      &   constant speed of light                        \\
$\mathbb{A}$                    &   $a_0^\prime = a_0$                      &   conservation law of gravitational              \\
                                &                                           &   ~~~~~scalar potential                          \\
$\mathbb{F}$                    &   $f_0^\prime = f_0$                      &   gauge equation, $f_0 = 0$                      \\
$\mathbb{P}$                    &   $m_g^\prime = m_g$                      &   mass conservation law                          \\
$\mathbb{L}$                    &   $L_{10}^\prime = L_{10}$                &   conservation law of the dot product of         \\
                                &                                           &   ~~~~~angular momenta                           \\
$\mathbb{W}$                    &   $W_{10}^{i \prime} = W_{10}^i$          &   energy conservation law                        \\
                                &   $W_{10}^\prime = W_{10}$                &   conservation law of the divergence of          \\
                                &                                           &   ~~~~~angular momenta                           \\
$\mathbb{N}$                    &   $N_{10}^{i \prime} = N_{10}^i$          &   conservation law of torque divergence          \\
                                &   $N_{10}^\prime = N_{10}$                &   power conservation law                         \\
\noalign{\smallskip}\hline
\end{tabular}
\end{table}

\section{\label{sec:level1}Electromagnetic fields in vacuum}

In the octonion space, if you multiply the basis vector $\textbf{I}_0$ by the octonion physical quantities, $\mathbb{R}$ , $\mathbb{V}$ , $\mathbb{A}$ , $\mathbb{F}$ , $\mathbb{P}$ , $\mathbb{L}$ , $\mathbb{W}$ , and $\mathbb{N}$ , from the left, you can get the new physical quantities, $\mathbb{R}_u$ , $\mathbb{V}_u$ , $\mathbb{A}_u$ , $\mathbb{F}_u$ , $\mathbb{P}_u$ , $\mathbb{L}_u$ , $\mathbb{W}_u$ , and $\mathbb{N}_u$ , according to the multiplication of octonions. Therefore the octonion physical quantities can be transformed from one octonion space, $\mathbb{O} ( r_j , R_j )$ , to another octonion space, $\mathbb{O}_u ( R_j , r_j )$ . Obviously, the two octonion spaces, $\mathbb{O}$ and $\mathbb{O}_u$ , are different from each other. In other words, the invariants in the octonion space, $\mathbb{O}_u$ , are distinct from these in the octonion space, $\mathbb{O}$ .

1) In the octonion space $\mathbb{O} ( r_j , R_j )$, the coordinate values of the physical quantities are related to the basis vector $\textbf{i}_j$ , in the gravitational fields. Meanwhile the coordinate values of the physical quantities are relevant to the basis vector $\textbf{I}_j$ , in the electromagnetic fields. By means of the properties of rotational transformations of octonion coordinate systems, it is able to deduce some invariants related to the basis vector $\textbf{i}_0$ , from the coordinate values of the physical quantities in the gravitational fields.

2) In another octonion space $\mathbb{O}_u ( R_j , r_j )$, the coordinate values of the physical quantities are related to the basis vector $\textbf{I}_j$ , in the gravitational fields. Meanwhile the coordinate values of the physical quantities are relevant to the basis vector $\textbf{i}_j$ , in the electromagnetic fields. The rotational transformations of octonion coordinate systems allow us to infer some invariants related to the basis vector $\textbf{i}_0$ , from the coordinate values of the physical quantities in the electromagnetic fields.

When the octonion coordinate systems are transformed rotationally, the above analysis reveals that the coordinate values of physical quantities of the electromagnetic fields must not be invariant, if those of the gravitational fields are invariants. Inversely, the coordinate values of physical quantities of the gravitational fields must not be invariant, in case those of the electromagnetic fields are invariants.

In a vacuum without considering the contribution of any material medium, when one octonion coordinate system, $\zeta$ , is transformed rotationally to another octonion coordinate system, $\eta$ , the physical quantities, $\mathbb{R}_u$ , $\mathbb{V}_u$ , $\mathbb{A}_u$ , $\mathbb{F}_u$, $\mathbb{P}_u$ , $\mathbb{L}_u$ , $\mathbb{W}_u$ , and $\mathbb{N}_u$ , can transform into the physical quantities, $\mathbb{R}_u^\prime$ , $\mathbb{V}_u^\prime$ , $\mathbb{A}_u^\prime$ , $\mathbb{F}_u^\prime$ , $\mathbb{P}_u^\prime$ , $\mathbb{L}_u^\prime$ , $\mathbb{W}_u^\prime$ , and $\mathbb{N}_u^\prime$ , respectively.

In the octonion coordinate system $\zeta$ , the octonion physical quantities, $\mathbb{R}_u$ , $\mathbb{V}_u$, $\mathbb{A}_u$ , $\mathbb{F}_u$, $\mathbb{P}_u$ , $\mathbb{L}_u$ , $\mathbb{W}_u$ , and $\mathbb{N}_u$ , can be written as follows,
\begin{eqnarray}
&& \mathbb{R}_u = k_{eg} ( - i R_0 + \Sigma \textbf{i}_k R_k ) + ( i \textbf{I}_0 r_0 - \Sigma \textbf{I}_k r_k ) ~ ,
\\
&& \mathbb{V}_u = k_{eg} ( - i V_0 + \Sigma \textbf{i}_k V_k ) + ( i \textbf{I}_0 v_0 - \Sigma \textbf{I}_k v_k ) ~ ,
\\
&& \mathbb{A}_u = k_{eg} ( - i A_0 + \Sigma \textbf{i}_k A_k ) + ( i \textbf{I}_0 a_0 - \Sigma \textbf{I}_k a_k ) ~ ,
\\
&& \mathbb{F}_u = k_{eg} ( - F_0 + \Sigma \textbf{i}_k F_k ) + ( \textbf{I}_0 f_0 - \Sigma \textbf{I}_k f_k ) ~ ,
\\
&& \mathbb{P}_u = k_{eg}^{-1} ( - i S_0 + \Sigma \textbf{i}_k S_k ) + ( i \textbf{I}_0 s_0 - \Sigma \textbf{I}_k s_k ) ~ ,
\\
&& \mathbb{L}_u = k_{eg} ( - L_{20} + i \Sigma \textbf{i}_k L_{2k}^i + \Sigma \textbf{i}_k L_{2k} )
\nonumber
\\
&&~~~~~~~
+ ( \textbf{I}_0 L_{10} - i \Sigma \textbf{I}_k L_{1k}^i - \Sigma \textbf{I}_k L_{1k} ) ~ ,
\\
&& \mathbb{W}_u = k_{eg} ( - i W_{20}^i - W_{20} + i \Sigma \textbf{i}_k W_{2k}^i + \Sigma \textbf{i}_k W_{2k} )
\nonumber
\\
&&~~~~~~~
+ ( i \textbf{I}_0 W_{10}^i + \textbf{I}_0 W_{10} - i \Sigma \textbf{I}_k W_{1k}^i - \Sigma \textbf{I}_k W_{1k} ) ~ ,
\\
&& \mathbb{N}_u = k_{eg} ( - i N_{20}^i - N_{20} + i \Sigma \textbf{i}_k N_{2k}^i + \Sigma \textbf{i}_k N_{2k} )
\nonumber
\\
&&~~~~~~~
+ ( i \textbf{I}_0 N_{10}^i + \textbf{I}_0 N_{10} - i \Sigma \textbf{I}_k N_{1k}^i - \Sigma \textbf{I}_k N_{1k} ) ~ .
\end{eqnarray}

In the octonion coordinate system $\eta$ , the octonion physical quantities, $\mathbb{R}_u^\prime$ , $\mathbb{V}_u^\prime$, $\mathbb{A}_u^\prime$ , $\mathbb{F}_u^\prime$, $\mathbb{P}_u^\prime$ , $\mathbb{L}_u^\prime$ , $\mathbb{W}_u^\prime$ , and $\mathbb{N}_u^\prime$ , can be written as, respectively,
\begin{eqnarray}
&& \mathbb{R}_u^\prime =  k_{eg} ( - i R_0^\prime + \Sigma \textbf{i}_k^\prime R_k^\prime ) + ( i \textbf{I}_0^\prime r_0^\prime - \Sigma \textbf{I}_k^\prime r_k^\prime ) ~ ,
\\
&& \mathbb{V}_u^\prime =  k_{eg} ( - i V_0^\prime + \Sigma \textbf{i}_k^\prime V_k^\prime ) + ( i \textbf{I}_0^\prime v_0^\prime - \Sigma \textbf{I}_k^\prime v_k^\prime ) ~ ,
\\
&& \mathbb{A}_u^\prime =  k_{eg} ( - i A_0^\prime + \Sigma \textbf{i}_k^\prime A_k^\prime ) + ( i \textbf{I}_0^\prime a_0^\prime - \Sigma \textbf{I}_k^\prime a_k^\prime ) ~ ,
\\
&& \mathbb{F}_u^\prime =  k_{eg} ( - F_0^\prime + \Sigma \textbf{i}_k^\prime F_k^\prime ) + ( \textbf{I}_0^\prime f_0^\prime - \Sigma \textbf{I}_k^\prime f_k^\prime ) ~ ,
\\
&& \mathbb{P}_u^\prime =  k_{eg}^{-1} ( - i S_0^\prime + \Sigma \textbf{i}_k^\prime S_k^\prime ) + ( i \textbf{I}_0^\prime s_0^\prime - \Sigma \textbf{I}_k^\prime s_k^\prime ) ~ ,
\\
&& \mathbb{L}_u^\prime =  k_{eg} ( - L_{20}^\prime + i \Sigma \textbf{i}_k^\prime L_{2k}^{i \prime} + \Sigma \textbf{i}_k^\prime L_{2k}^\prime )
\nonumber
\\
&&~~~~~~~
+ ( \textbf{I}_0^\prime L_{10}^\prime - i \Sigma \textbf{I}_k^\prime L_{1k}^{i \prime} - \Sigma \textbf{I}_k^\prime L_{1k}^\prime ) ~ ,
\\
&& \mathbb{W}_u^\prime =  k_{eg} ( - i W_{20}^{i \prime} - W_{20}^\prime + i \Sigma \textbf{i}_k^\prime W_{2k}^{i \prime} + \Sigma \textbf{i}_k^\prime W_{2k}^\prime )
\nonumber
\\
&&~~~~~~~
+ ( i \textbf{I}_0^\prime W_{10}^{i \prime} + \textbf{I}_0^\prime W_{10}^\prime - i \Sigma \textbf{I}_k^\prime W_{1k}^{i \prime} - \Sigma \textbf{I}_k^\prime W_{1k}^\prime ) ~ ,
\\
&& \mathbb{N}_u^\prime =  k_{eg} ( - i N_{20}^{i \prime} - N_{20}^\prime + i \Sigma \textbf{i}_k^\prime N_{2k}^{i \prime} + \Sigma \textbf{i}_k^\prime N_{2k}^\prime )
\nonumber
\\
&&~~~~~~~
+ ( i \textbf{I}_0^\prime N_{10}^{i \prime} + \textbf{I}_0^\prime N_{10}^\prime - i \Sigma \textbf{I}_k^\prime N_{1k}^{i \prime} - \Sigma \textbf{I}_k^\prime N_{1k}^\prime ) ~ .
\end{eqnarray}

When the octonion physical quantities are transformed rotationally from one octonion coordinate system, $\zeta$ , to another octonion coordinate system, $\eta$ , each of scalar parts of octonion physical quantities, relevant to electromagnetic fields, is invariant under the coordinate transformations. From Eqs.(33)-(48), there are ten invariants as follows,
\begin{eqnarray}
&& R_0^\prime = R_0 ~ ,
\\
&& V_0^\prime = V_0 ~ ,
\\
&& A_0^\prime = A_0 ~ ,
\\
&& F_0^\prime = F_0 ~ ,
\\
&& S_0^\prime = S_0 ~ ,
\\
&& L_{20}^\prime = L_{20} ~ ,
\\
&& W_{20}^{i \prime} = W_{20}^i ~ ,
\\
&& W_{20}^\prime = W_{20} ~ ,
\\
&& N_{20}^{i \prime} = N_{20}^i ~ ,
\\
&& N_{20}^\prime = N_{20} ~ .
\end{eqnarray}

From Eqs.(50) and (53), it is able to achieve the charge conservation law. From Eq.(55), it will deduce the second-energy conservation law. From Eq.(58), it can infer the second-power conservation law. Eqs.(51), (52), and (54) show that the scalar part of electromagnetic potential, the scalar part of electromagnetic strength, and the dot product of magnetic moments are the invariants independent of each other. Eqs.(56) and (57) state that the divergence of magnetic moments and the divergence of second-torques are invariants also.

The above analysis shows that there are multiple conservation laws, in the octonion space $\mathbb{O}_u$ . And that these conservation laws are able to be effective simultaneously. For instance, the charge conservation law, second-energy conservation law, and second-power conservation law can be valid simultaneously (Table 2).

However, the laws of conservation in two different types of octonion spaces, $\mathbb{O}$ and $\mathbb{O}_u$ , are unable to be effective simultaneously. a) The rotational transformations of coordinate systems, in the octonion space $\mathbb{O}$ , are able to deduce the mass conservation law. Meanwhile the term related to electric-charge is a vector in the octonion space $\mathbb{O}$ , and that this vector will vary in the rotational transformations in the octonion space $\mathbb{O}$ . b) Conversely, the rotational transformations of coordinate systems, in the octonion space $\mathbb{O}_u$ , are capable of inferring the charge conservation law. Meanwhile the term related to mass is a vector in the octonion space $\mathbb{O}_u$ , further this vector will change in the rotational transformations in the octonion space $\mathbb{O}_u$ .

The octonion space $\mathbb{O}_u$ is distinct from the octonion space $\mathbb{O}$ . As a result, the invariants in two different types of octonion spaces, $\mathbb{O}$ and $\mathbb{O}_u$ , are incompatible. That is, two different types of invariants are unable to be valid simultaneously, in the Tables 1 and 2. In particular, the charge conservation law and mass conservation law are not effective simultaneously. Similarly, the charge conservation law and energy conservation law are not valid simultaneously.

The above method can be extended from the gravitational and electromagnetic fields, in a vacuum without considering the contribution of any material medium, to those with considering the contribution of material media.

\begin{table}[h]
\centering
\caption{Some invariants and conservation laws relevant to electromagnetic fields, in the octonion space $\mathbb{O}_u$ without considering the contribution of any material medium.}
\label{tab:2}       
\begin{tabular}{@{}lll@{}}
\hline\noalign{\smallskip}
physical quantity               &   invariants                              &   conservation laws                               \\
\noalign{\smallskip}\hline\noalign{\smallskip}
$\mathbb{R}_u$                  &   $R_0^\prime = R_0$                      &   conservation law of the second-scalar part of   \\
                                &                                           &   ~~~~~octonion radius vector                     \\
$\mathbb{V}_u$                  &   $V_0^\prime = V_0$                      &   constant second-speed of light                  \\
$\mathbb{A}_u$                  &   $A_0^\prime = A_0$                      &   conservation law of electromagnetic             \\
                                &                                           &   ~~~~~scalar potential                           \\
$\mathbb{F}_u$                  &   $F_0^\prime = F_0$                      &   gauge equation, $F_0 = 0$                       \\
$\mathbb{P}_u$                  &   $q^\prime = q$                          &   charge conservation law                         \\
$\mathbb{L}_u$                  &   $L_{20}^\prime = L_{20}$                &   conservation law of the dot product of          \\
                                &                                           &   ~~~~~magnetic moments                           \\
$\mathbb{W}_u$                  &   $W_{20}^{i \prime} = W_{20}^i$          &   second-energy conservation law                  \\
                                &   $W_{20}^\prime = W_{20}$                &   conservation law of the divergence of           \\
                                &                                           &   ~~~~~magnetic moments                           \\
$\mathbb{N}_u$                  &   $N_{20}^{i \prime} = N_{20}^i$          &   conservation law of second-torque divergence    \\
                                &   $N_{20}^\prime = N_{20}$                &   second-power conservation law                   \\
\noalign{\smallskip}\hline
\end{tabular}
\end{table}

\section{\label{sec:level1}Field equations within material media}

In the octonion space $\mathbb{O}$ , the octonion field strength $\mathbb{F}$ and octonion angular momentum $\mathbb{L}$ can be combined together to become the octonion composite field strength, $\mathbb{F}^+ = \mathbb{F} + k_{fl} \mathbb{L}$. Herein $k_{fl} = - \mu_g$ , is one coefficient, to meet the need for dimensional homogeneity. That is, the octonion composite field strength, $\mathbb{F}^+$ , is the octonion field strength within material media for the gravitational and electromagnetic fields.

The octonion composite field source within material media,
\begin{eqnarray}
&&  \mu \mathbb{S}^+ = - ( i \mathbb{F}^+ / v_0 + \lozenge )^\ast \circ \mathbb{F}^+ ~ ,
\end{eqnarray}
can be rewritten as,
\begin{eqnarray}
&&  - ( i \mathbb{F}^+ / v_0 + \lozenge )^\ast \circ \mathbb{F}^+ = \mu_g \mathbb{S}_g^+ + k_{eg} \mu_e \mathbb{S}_e^+ - ( i \mathbb{F}^+ / v_0 )^\ast \circ \mathbb{F}^+  ~ ,
\end{eqnarray}
where $\mu \mathbb{S}^+ = \mu \mathbb{S} + k_{fl} \mathbb{Z}$ . The octonion physical quantity $\mathbb{Z}$ is the field source part only related to the material media. $\mathbb{Z} = \mathbb{Z}_g + k_{eg} \mathbb{Z}_e$ . $\mathbb{Z}_g$ and $\mathbb{Z}_e$ are the components of the octonion physical quantity $\mathbb{Z}$ in the two subspaces, $\mathbb{H}_g$ and $\mathbb{H}_{em}$ , respectively.

From the above, there are, $\mu_g \mathbb{S}_g^+ = - \lozenge^\ast \circ \mathbb{F}_g^+$ , and $\mu_e \mathbb{S}_e^+ = - \lozenge^\ast \circ \mathbb{F}_e^+$ . The former is the gravitational equations for the material media in the quaternion space $\mathbb{H}_g$, while the latter is the electromagnetic equations for the material media in the second subspace $\mathbb{H}_{em}$ . $\mu_g \mathbb{S}_g^+ = \mu_g \mathbb{S}_g + k_{fl} \mathbb{Z}_g$ . $\mu_e \mathbb{S}_e^+ = \mu_e \mathbb{S}_e + k_{fl} \mathbb{Z}_e$. $\mathbb{F}^+ = \mathbb{F}_g^+ + k_{eg} \mathbb{F}_e^+$. $\mathbb{F}_g^+$ and $\mathbb{F}_e^+$ are the components of the octonion composite field strength $\mathbb{F}^+$ in the two subspaces, $\mathbb{H}_g$ and $\mathbb{H}_{em}$ , respectively. For one single particle, $s_0^+ = m_g^+ v_0$ , with $m_g^+$ being the gravitational mass within material media. $S_0^+ = q^+ V_0$ , and $q^+$ is the electric charge within material media.

The octonion composite linear momentum, $\mathbb{P}^+$ , within material media can be written as,
\begin{eqnarray}
&&  \mathbb{P}^+ = \mu \mathbb{S}^+ / \mu_g  ~ ,
\end{eqnarray}
where the above can be rewritten as, $\mathbb{P}^+ = \mathbb{P}_g^+ + k_{eg} \mathbb{P}_e^+$ . $\mathbb{P}_g^+ = \{ \mu_g \mathbb{S}_g^+ - ( i \mathbb{F}^+ / v_0 )^\ast \circ \mathbb{F}^+ \} / \mu_g$, $\mathbb{P}_e^+ = \mu_e \mathbb{S}_e^+ / \mu_g$ . $\mathbb{P}_g^+$ and $\mathbb{P}_e^+$ are the components of the octonion composite linear momentum, $\mathbb{P}^+$ , for the material media in the two subspaces, $\mathbb{H}_g$ and $\mathbb{H}_{em}$, respectively. $\mathbb{P}_g^+ = i p_0^+ + \textbf{p}^+$ . $\textbf{p}^+ = \Sigma p_k^+ \textbf{i}_k$ . $\mathbb{P}_e^+ = i \textbf{P}_0^+ + \textbf{P}^+$ . $\textbf{P}^+ = \Sigma P_k^+ \textbf{I}_k$ . $\textbf{P}_0^+ = P_0^+ I_0$.

The octonion composite angular momentum, $\mathbb{L}^+$ , within material media can be written as,
\begin{eqnarray}
&&  \mathbb{L}^+ = ( \mathbb{R} + k_{rx} \mathbb{X} )^\times \circ \mathbb{P}^+  ~ ,
\end{eqnarray}
where the above can be rewritten as, $\mathbb{L}^+ = \mathbb{L}_g^+ + k_{eg} \mathbb{L}_e^+$ . $\mathbb{L}_g^+ = L_{10}^+ + i \textbf{L}_1^{i +} + \textbf{L}_1^+$. $\mathbb{L}_e^+ = \textbf{L}_{20}^+ + i \textbf{L}_2^{i +} + \textbf{L}_2^+$ . $\mathbb{L}_g^+$ and $\mathbb{L}_e^+$ are the components of the octonion composite angular momentum, $\mathbb{L}^+$ , for the material media in the two subspaces, $\mathbb{H}_g$ and $\mathbb{H}_{em}$ , respectively. $\textbf{L}_1^+$ is the angular momentum within material media. $\textbf{L}_1^{i +}$ is the mass moment within material media. $\textbf{L}_2^{i +}$ is the electric moment within material media, and $\textbf{L}_2^+$ is the magnetic moment within material media. $\textbf{L}_1^+ = \Sigma L_{1k}^+ \textbf{i}_k$ , $\textbf{L}_1^{i +} = \Sigma L_{1k}^{i +} \textbf{i}_k$ . $\textbf{L}_2^+ = \Sigma L_{2k}^+ \textbf{I}_k$ , $\textbf{L}_2^{i +} = \Sigma L_{2k}^{i +} \textbf{I}_k$ . $\textbf{L}_{20}^+ = \textbf{I}_0 L_{20}^+$ . $L_{1j}^+ , L_{2j}^+ , L_{1k}^{i +}$ , and $L_{2k}^{i +}$ are all real.

The octonion composite torque, $\mathbb{W}^+$ , within material media can be written as,
\begin{eqnarray}
&&  \mathbb{W}^+ = - v_0 ( i \mathbb{F}^+ / v_0 + \lozenge ) \circ \{ ( i \mathbb{V}^\times / v_0 ) \circ \mathbb{L}^+ \} ~ ,
\end{eqnarray}
where the above can be rewritten as, $\mathbb{W}^+ = \mathbb{W}_g^+ + k_{eg} \mathbb{W}_e^+$ . $\mathbb{W}_g^+ = i W_{10}^{i +} + W_{10}^+ + i \textbf{W}_1^{i +} + \textbf{W}_1^+$. $\mathbb{W}_e^+ = i \textbf{W}_{20}^{i +} + \textbf{W}_{20}^+ + i \textbf{W}_2^{i +} + \textbf{W}_2^+$ . $\mathbb{W}_g^+$ and $\mathbb{W}_e^+$ are the components of the octonion composite torque, $\mathbb{W}^+$ , for the material media in the two subspaces, $\mathbb{H}_g$ and $\mathbb{H}_{em}$ , respectively. $W_{10}^{i +}$ is the energy within material media. $\textbf{W}_1^{i +}$ is the torque within material media, including the gyroscopic torque, $\nabla ( \textbf{v} \cdot \textbf{L}_1^+ )$, within material media. $\textbf{W}_{20}^{i +}$ and $\textbf{W}_2^{i +}$ are the second energy and second torque within material media, respectively. $\textbf{W}_1^+ = \Sigma W_{1k}^+ \textbf{i}_k$, $\textbf{W}_1^{i +} = \Sigma W_{1k}^{i +} \textbf{i}_k$. $\textbf{W}_2^+ = \Sigma W_{2k}^+ \textbf{I}_k$ , $\textbf{W}_2^{i +} = \Sigma W_{2k}^{i +} \textbf{I}_k$ , $\textbf{W}_{20}^{i +} = W_{20}^{i +} \textbf{I}_0$ . $\textbf{W}_{20}^+ = W_{20}^+ \textbf{I}_0$ . $W_{1j}^+ , W_{2j}^+ , W_{1j}^{i +}$, and $W_{2j}^{i +}$ are all real.

The octonion composite force, $\mathbb{N}^+$ , within material media can be written as,
\begin{eqnarray}
&&  \mathbb{N}^+ = - ( i \mathbb{F}^+ / v_0 + \lozenge ) \circ \{ ( i \mathbb{V}^\times / v_0 ) \circ \mathbb{W}^+ \} ~ ,
\end{eqnarray}
where the above can be rewritten as, $\mathbb{N}^+ = \mathbb{N}_g^+ + k_{eg} \mathbb{N}_e^+$ . $\mathbb{N}_g^+ = i N_{10}^{i +} + N_{10}^+ + i \textbf{N}_1^{i +} + \textbf{N}_1^+$ . $\mathbb{N}_e^+ = i \textbf{N}_{20}^{i +} + \textbf{N}_{20}^+ + i \textbf{N}_2^{i +} + \textbf{N}_2^+$. $\mathbb{N}_g^+$ and $\mathbb{N}_e^+$ are the components of the octonion composite force, $\mathbb{N}^+$ , for the material media in the two subspaces, $\mathbb{H}_g$ and $\mathbb{H}_{em}$, respectively. $N_{10}^+$ is the power within material media. $\textbf{N}_1^{i +}$ is the force within material media, including the Magnus force, $ \nabla ( \partial L_{10}^+ / \partial t)$ , within material media. $\textbf{N}_{20}^+$ is the second power within material media. $\textbf{N}_2^{i +}$ is the second force within material media. $\textbf{N}_1^+ = \Sigma N_{1k}^+ \textbf{i}_k$ , $\textbf{N}_1^{i +} = \Sigma N_{1k}^{i +} \textbf{i}_k$. $\textbf{N}_2^+ = \Sigma N_{2k}^+ \textbf{I}_k$, $\textbf{N}_2^{i +} = \Sigma N_{2k}^{i +} \textbf{I}_k$ , $\textbf{N}_{20}^{i +} = N_{20}^{i +} \textbf{I}_0$ . $\textbf{N}_{20}^+ = N_{20} \textbf{I}_0^+$ . $N_{1j}^+ , N_{2j}^+ , N_{1j}^{i +}$ , and $N_{2j}^{i +}$ are all real.

What is noteworthy is that the octonion composite field strength, linear momentum, angular momentum, torque, and force involve with the material media, in the octonion spaces. The five physical quantities within material media are different from their cases in vacuum, respectively. However, each of the octonion radius vector, field potential, and integrating function of field potential does not concern any material medium. The three physical quantities remain unchanged, within any material medium.

\begin{table}[h]
\centering
\caption{Some invariants and conservation laws, relevant to gravitational fields within material media, in the octonion space $\mathbb{O}$ .}
\label{tab:2}       
\begin{tabular}{@{}lll@{}}
\hline\noalign{\smallskip}
physical quantity               &   invariants                                   &   conservation laws                               \\
\noalign{\smallskip}\hline\noalign{\smallskip}
$\mathbb{F}^+$                  &   $f_0^{+ \prime} = f_0^+$                     &   gauge equation, $f_0^+ = 0$                     \\
$\mathbb{P}^+$                  &   $m_g^{+ \prime} = m_g^+$                     &   mass conservation law                           \\
$\mathbb{L}^+$                  &   $L_{10}^{+ \prime} = L_{10}^+$               &   conservation law of the dot product of          \\
                                &                                                &   ~~~~~angular momenta                            \\
$\mathbb{W}^+$                  &   $W_{10}^{i + \prime} = W_{10}^{i +}$         &   energy conservation law                         \\
                                &   $W_{10}^{+ \prime} = W_{10}^+$               &   conservation law of the divergence of           \\
                                &                                                &   ~~~~~angular momenta                            \\
$\mathbb{N}^+$                  &   $N_{10}^{i + \prime} = N_{10}^{i +}$         &   conservation law of torque divergence           \\
                                &   $N_{10}^{+ \prime} = N_{10}^+$               &   power conservation law                          \\
\noalign{\smallskip}\hline
\end{tabular}
\end{table}

\section{\label{sec:level1}Invariants within material media}

In the octonion spaces, the above method of invariants can be extended from the gravitational and electromagnetic fields in vacuum to these within material media, under rotational transformations of octonion systems.

\subsection{\label{sec:level1}Gravitational fields within material media}

In the octonion spaces where the contribution of the material media is significant enough, when an octonion coordinate system, $\alpha$ , is transformed rotationally to another octonion coordinate system, $\beta$ , the physical quantities, $\mathbb{F}^+$ , $\mathbb{P}^+$ , $\mathbb{L}^+$ , $\mathbb{W}^+$, and $\mathbb{N}^+$ , can transform into the physical quantities, $\mathbb{F}^{+ \prime}$ , $\mathbb{P}^{+ \prime}$ , $\mathbb{L}^{+ \prime}$ , $\mathbb{W}^{+ \prime}$ , and $\mathbb{N}^{+ \prime}$ , respectively.

In the octonion coordinate system $\alpha$ , the octonion physical quantities, $\mathbb{F}^+$ , $\mathbb{P}^+$, $\mathbb{L}^+$ , $\mathbb{W}^+$ , and $\mathbb{N}^+$ ,  within material media can be written as follows,
\begin{eqnarray}
&& \mathbb{F}^+ = f_0^+ + \Sigma \textbf{i}_k f_k^+ + k_{eg} ( \textbf{I}_0 F_0^+ + \Sigma \textbf{I}_k F_k^+ ) ~ ,
\\
&& \mathbb{P}^+ = i s_0^+ + \Sigma \textbf{i}_k s_k^+ + k_{eg}^{-1} ( i \textbf{I}_0 S_0^+ + \Sigma \textbf{I}_k S_k^+ ) ~ ,
\\
&& \mathbb{L}^+ = L_{10}^+ + i \Sigma \textbf{i}_k L_{1k}^{i +} + \Sigma \textbf{i}_k L_{1k}^+
\nonumber
\\
&&~~~~~~~
+ k_{eg} ( \textbf{I}_0 L_{20}^+ + i \Sigma \textbf{I}_k L_{2k}^{i +} + \Sigma \textbf{I}_k L_{2k}^+  ) ~ ,
\\
&& \mathbb{W}^+ = i W_{10}^{i +} + W_{10}^+ + i \Sigma \textbf{i}_k W_{1k}^{i +} + \Sigma \textbf{i}_k W_{1k}^+
\nonumber
\\
&&~~~~~~~
+ k_{eg} ( i \textbf{I}_0 W_{20}^{i +} + \textbf{I}_0 W_{20}^+ + i \Sigma \textbf{I}_k W_{2k}^{i +} + \Sigma \textbf{I}_k W_{2k}^+ ) ~ ,
\\
&& \mathbb{N}^+ = i N_{10}^{i +} + N_{10}^+ + i \Sigma \textbf{i}_k N_{1k}^{i +} + \Sigma \textbf{i}_k N_{1k}^+
\nonumber
\\
&&~~~~~~~
+ k_{eg} ( i \textbf{I}_0 N_{20}^{i +} + \textbf{I}_0 N_{20}^+ + i \Sigma \textbf{I}_k N_{2k}^{i +} + \Sigma \textbf{I}_k N_{2k}^+ ) ~ .
\end{eqnarray}

In the octonion coordinate system $\beta$ , the octonion physical quantities, $\mathbb{F}^{+ \prime}$ , $\mathbb{P}^{+ \prime}$ , $\mathbb{L}^{+ \prime}$ , $\mathbb{W}^{+ \prime}$ , and $\mathbb{N}^{+ \prime}$ , within material media can be written as, respectively,
\begin{eqnarray}
&& \mathbb{F}^{+ \prime} = f_0^{+ \prime} + \Sigma \textbf{i}_k^\prime f_k^{+ \prime} + k_{eg} ( \textbf{I}_0^\prime F_0^{+ \prime} + \Sigma \textbf{I}_k^\prime F_k^{+ \prime} ) ~ ,
\\
&& \mathbb{P}^{+ \prime} = i s_0^{+ \prime} + \Sigma \textbf{i}_k^\prime s_k^{+ \prime} + k_{eg}^{-1} ( i \textbf{I}_0^\prime S_0^{+ \prime} + \Sigma \textbf{I}_k^\prime S_k^{+ \prime} ) ~ ,
\\
&& \mathbb{L}^{+ \prime} = L_{10}^{+ \prime} + i \Sigma \textbf{i}_k^\prime L_{1k}^{i + \prime} + \Sigma \textbf{i}_k^\prime L_{1k}^{+ \prime}
\nonumber
\\
&&~~~~~~~
+ k_{eg} ( \textbf{I}_0^\prime L_{20}^{+ \prime} + i \Sigma \textbf{I}_k^\prime L_{2k}^{i + \prime} + \Sigma \textbf{I}_k^\prime L_{2k}^{+ \prime}  ) ~ ,
\\
&& \mathbb{W}^{+ \prime} = i W_{10}^{i + \prime} + W_{10}^{+ \prime} + i \Sigma \textbf{i}_k^\prime W_{1k}^{i + \prime} + \Sigma \textbf{i}_k^\prime W_{1k}^{+ \prime}
\nonumber
\\
&&~~~~~~~
+ k_{eg} ( i \textbf{I}_0^\prime W_{20}^{i + \prime} + \textbf{I}_0^\prime W_{20}^{+ \prime} + i \Sigma \textbf{I}_k^\prime W_{2k}^{i + \prime} + \Sigma \textbf{I}_k^\prime W_{2k}^{+ \prime} ) ~ ,
\\
&& \mathbb{N}^{+ \prime} = i N_{10}^{i + \prime} + N_{10}^{+ \prime} + i \Sigma \textbf{i}_k^\prime N_{1k}^{i + \prime} + \Sigma \textbf{i}_k^\prime N_{1k}^{+ \prime}
\nonumber
\\
&&~~~~~~~
+ k_{eg} ( i \textbf{I}_0^\prime N_{20}^{i + \prime} + \textbf{I}_0^\prime N_{20}^{+ \prime} + i \Sigma \textbf{I}_k^\prime N_{2k}^{i + \prime} + \Sigma \textbf{I}_k^\prime N_{2k}^{+ \prime} ) ~ .
\end{eqnarray}

When the octonion physical quantities within material media are transformed rotationally from the octonion coordinate system, $\alpha$ , to another octonion coordinate system, $\beta$ , each of scalar parts of octonion physical quantities, relevant to gravitational fields within material media, is invariant under the coordinate transformations. From Eqs.(65)-(74), there are seven invariants as follows,
\begin{eqnarray}
&& f_0^{+ \prime} = f_0^+ ~ ,
\\
&& s_0^{+ \prime} = s_0^+ ~ ,
\\
&& L_{10}^{+ \prime} = L_{10}^+ ~ ,
\\
&& W_{10}^{i + \prime} = W_{10}^{i +} ~ ,
\\
&& W_{10}^{+ \prime} = W_{10}^+ ~ ,
\\
&& N_{10}^{i + \prime} = N_{10}^{i +} ~ ,
\\
&& N_{10}^{+ \prime} = N_{10}^+ ~ .
\end{eqnarray}

From Eqs.(6) and (76), it is able to achieve the mass conservation law within material media. From Eq.(78), it will deduce the energy conservation law within material media. From Eq.(81), it can infer the power conservation law within material media. Eqs.(75) and (77) imply that the scalar part of gravitational strength, as well as the dot product of angular momenta, are the invariants independent of each other within material media. Eqs.(79) and (80) state that the divergence of angular momenta and the divergence of torques are invariants within material media also.

The above analysis shows that there are multiple conservation laws within material media, in the octonion space $\mathbb{O}$. And that these conservation laws within material media are able to be effective simultaneously. For instance, the mass conservation law, energy conservation law, and power conservation law can be established within material media simultaneously (Table 3).

The above method can be extended from the gravitational fields within material media to the electromagnetic fields within material media.

\begin{table}[h]
\centering
\caption{Some invariants and conservation laws, relevant to electromagnetic fields within material media, in the octonion space $\mathbb{O}_u$ .}
\label{tab:2}       
\begin{tabular}{@{}lll@{}}
\hline\noalign{\smallskip}
physical quantity               &   invariants                                   &   conservation laws                             \\
\noalign{\smallskip}\hline\noalign{\smallskip}
$\mathbb{F}_u^+$                &   $F_0^{+ \prime} = F_0^+$                     &   gauge equation, $F_0^+ = 0$                   \\
$\mathbb{P}_u^+$                &   $q^{+ \prime} = q^+$                         &   charge conservation law                       \\
$\mathbb{L}_u^+$                &   $L_{20}^{+ \prime} = L_{20}^+$               &   conservation law of the dot product of        \\
                                &                                                &   ~~~~~magnetic moments                         \\
$\mathbb{W}_u^+$                &   $W_{20}^{i + \prime} = W_{20}^{i +}$         &   second-energy conservation law                \\
                                &   $W_{20}^{+ \prime} = W_{20}^+$               &   conservation law of the divergence of         \\
                                &                                                &   ~~~~~magnetic moments                         \\
$\mathbb{N}_u^+$                &   $N_{20}^{i + \prime} = N_{20}^{i +}$         &   conservation law of second-torque divergence  \\
                                &   $N_{20}^{+ \prime} = N_{20}^+$               &   second-power conservation law                 \\
\noalign{\smallskip}\hline
\end{tabular}
\end{table}

\subsection{\label{sec:level1}Electromagnetic fields within material media}

In the octonion spaces where the contribution of the material media is significant enough, the octonion physical quantities, $\mathbb{F}_u^+$ , $\mathbb{P}_u^+$ , $\mathbb{L}_u^+$ , $\mathbb{W}_u^+$, and $\mathbb{N}_u^+$ , can transform into the octonion physical quantities, $\mathbb{F}_u^{+ \prime}$ , $\mathbb{P}_u^{+ \prime}$ , $\mathbb{L}_u^{+ \prime}$ , $\mathbb{W}_u^{+ \prime}$ , and $\mathbb{N}_u^{+ \prime}$ , respectively, when an octonion coordinate system, $\zeta$ , is transformed rotationally to another octonion coordinate system, $\eta$ .

In the octonion coordinate system $\zeta$ , the octonion physical quantities, $\mathbb{F}_u^+$ , $\mathbb{P}_u^+$, $\mathbb{L}_u^+$ , $\mathbb{W}_u^+$ , and $\mathbb{N}_u^+$ , within material media can be written as follows,
\begin{eqnarray}
&& \mathbb{F}_u^+ = k_{eg} ( - F_0^+ + \Sigma \textbf{i}_k F_k^+ ) + ( \textbf{I}_0 f_0^+ - \Sigma \textbf{I}_k f_k^+ ) ~ ,
\\
&& \mathbb{P}_u^+ = k_{eg}^{-1} ( - i S_0^+ + \Sigma \textbf{i}_k S_k^+ ) + ( i \textbf{I}_0 s_0^+ - \Sigma \textbf{I}_k s_k^+ ) ~ ,
\\
&& \mathbb{L}_u^+ = k_{eg} ( - L_{20}^+ + i \Sigma \textbf{i}_k L_{2k}^{i +} + \Sigma \textbf{i}_k L_{2k}^+ )
\nonumber
\\
&&~~~~~~~
+ ( \textbf{I}_0 L_{10}^+ - i \Sigma \textbf{I}_k L_{1k}^{i +} - \Sigma \textbf{I}_k L_{1k}^+ ) ~ ,
\\
&& \mathbb{W}_u^+ = k_{eg} ( - i W_{20}^{i +} - W_{20}^+ + i \Sigma \textbf{i}_k W_{2k}^{i +} + \Sigma \textbf{i}_k W_{2k}^+ )
\nonumber
\\
&&~~~~~~~
+ ( i \textbf{I}_0 W_{10}^{i +} + \textbf{I}_0 W_{10}^+ - i \Sigma \textbf{I}_k W_{1k}^{i +} - \Sigma \textbf{I}_k W_{1k}^+ ) ~ ,
\\
&& \mathbb{N}_u^+ = k_{eg} ( - i N_{20}^{i +} - N_{20}^+ + i \Sigma \textbf{i}_k N_{2k}^{i +} + \Sigma \textbf{i}_k N_{2k}^+ )
\nonumber
\\
&&~~~~~~~
+ ( i \textbf{I}_0 N_{10}^{i +} + \textbf{I}_0 N_{10}^+ - i \Sigma \textbf{I}_k N_{1k}^{i +} - \Sigma \textbf{I}_k N_{1k}^+ ) ~ .
\end{eqnarray}

In the octonion coordinate system $\eta$ , the octonion physical quantities, $\mathbb{F}_u^{+ \prime}$ , $\mathbb{P}_u^{+ \prime}$, $\mathbb{L}_u^{+ \prime}$ , $\mathbb{W}_u^{+ \prime}$ , and $\mathbb{N}_u^{+ \prime}$ , within material media can be written as, respectively,
\begin{eqnarray}
&& \mathbb{F}_u^{+ \prime} = k_{eg} ( - F_0^{+ \prime} + \Sigma \textbf{i}_k^\prime F_k^{+ \prime} ) + ( \textbf{I}_0^\prime f_0^{+ \prime} - \Sigma \textbf{I}_k^\prime f_k^{+ \prime} ) ~ ,
\\
&& \mathbb{P}_u^{+ \prime} = k_{eg}^{-1} ( - i S_0^{+ \prime} + \Sigma \textbf{i}_k^\prime S_k^{+ \prime} ) + ( i \textbf{I}_0^\prime s_0^{+ \prime} - \Sigma \textbf{I}_k^\prime s_k^{+ \prime} ) ~ ,
\\
&& \mathbb{L}_u^{+ \prime} = k_{eg} ( - L_{20}^{+ \prime} + i \Sigma \textbf{i}_k^\prime L_{2k}^{i + \prime} + \Sigma \textbf{i}_k^\prime L_{2k}^{+ \prime} )
\nonumber
\\
&&~~~~~~~
+ ( \textbf{I}_0^\prime L_{10}^{+ \prime} - i \Sigma \textbf{I}_k^\prime L_{1k}^{i + \prime} - \Sigma \textbf{I}_k^\prime L_{1k}^{+ \prime} ) ~ ,
\\
&& \mathbb{W}_u^{+ \prime} = k_{eg} ( - i W_{20}^{i + \prime} - W_{20}^{+ \prime} + i \Sigma \textbf{i}_k^\prime W_{2k}^{i + \prime} + \Sigma \textbf{i}_k^\prime W_{2k}^{+ \prime} )
\nonumber
\\
&&~~~~~~~
+ ( i \textbf{I}_0^\prime W_{10}^{i + \prime} + \textbf{I}_0^\prime W_{10}^{+ \prime} - i \Sigma \textbf{I}_k^\prime W_{1k}^{i + \prime} - \Sigma \textbf{I}_k^\prime W_{1k}^{+ \prime} ) ~ ,
\\
&& \mathbb{N}_u^{+ \prime} = k_{eg} ( - i N_{20}^{i + \prime} - N_{20}^{+ \prime} + i \Sigma \textbf{i}_k^\prime N_{2k}^{i + \prime} + \Sigma \textbf{i}_k^\prime N_{2k}^{+ \prime} )
\nonumber
\\
&&~~~~~~~
+ ( i \textbf{I}_0^\prime N_{10}^{i + \prime} + \textbf{I}_0^\prime N_{10}^{+ \prime} - i \Sigma \textbf{I}_k^\prime N_{1k}^{i + \prime} - \Sigma \textbf{I}_k^\prime N_{1k}^{+ \prime} ) ~ .
\end{eqnarray}

When the octonion physical quantities within material media are transformed rotationally from the octonion coordinate system, $\zeta$ , to another octonion coordinate system, $\eta$ , each of scalar parts of octonion physical quantities, relevant to electromagnetic fields within material media, is invariant under the coordinate transformations. From Eqs.(82)-(91), there are seven invariants as follows,
\begin{eqnarray}
&& F_0^{+ \prime} = F_0^+ ~ ,
\\
&& S_0^{+ \prime} = S_0^+ ~ ,
\\
&& L_{20}^{+ \prime} = L_{20}^+ ~ ,
\\
&& W_{20}^{i + \prime} = W_{20}^{i +} ~ ,
\\
&& W_{20}^{+ \prime} = W_{20}^+ ~ ,
\\
&& N_{20}^{i + \prime} = N_{20}^{i +} ~ ,
\\
&& N_{20}^{+ \prime} = N_{20}^+ ~ .
\end{eqnarray}

From Eqs.(50) and (93), it is able to achieve the charge conservation law within material media. From Eq.(95), it will deduce the second-energy conservation law within material media. From Eq.(98), it can infer the second-power conservation law within material media. Eqs.(92) and (94) imply that the scalar part of electromagnetic strength, as well as the dot product of magnetic moments, are the invariants independent of each other within material media. Eqs.(96) and (97) show that the divergence of magnetic moments and the divergence of second-torques are invariants within material media.

The above analysis shows that there are multiple conservation laws within material media, in another octonion space $\mathbb{O}_u$ . And that these conservation laws within material media are able to be effective simultaneously. For instance, the charge conservation law, second-energy conservation law, and second-power conservation law can be valid within material media simultaneously.

However, the conservation laws in two different types of octonion spaces, $\mathbb{O}$ and $\mathbb{O}_u$ , are unable to be effective within material media simultaneously. a) The rotational transformations of coordinate systems, in the octonion space $\mathbb{O}$ , allow us to deduce the mass conservation law within material media. Meanwhile the term related to electric-charge within material media is a vector in the octonion space $\mathbb{O}$ , further this vector will alter in the rotational transformations in the octonion space $\mathbb{O}$ . b) Conversely, the rotational transformations of coordinate systems, in the octonion space $\mathbb{O}_u$ , are capable of inferring the charge conservation law within material media. Meanwhile the term related to mass within material media is a vector in the octonion space $\mathbb{O}_u$ , and that this vector will vary in the rotational transformations in the octonion space $\mathbb{O}_u$ .

The octonion space $\mathbb{O}_u$ is distinct from the octonion space $\mathbb{O}$ . As a result, the invariants in two different types of octonion spaces, $\mathbb{O}$ and $\mathbb{O}_u$ , are incompatible within material media. That is, two different types of invariants within material media are unable to be valid simultaneously, in the Tables 3 and 4. Strictly speaking, the charge conservation law and the mass conservation law can not be effective within material media simultaneously. Similarly, the charge conservation law and the energy conservation law are incompatible within material media (Table 4).

\section{\label{sec:level1}Conclusions and discussions}

The algebra of octonions can be applied to describe the physical properties of electromagnetic and gravitational fields, including the octonion field potential, field strength, field source, linear momentum, angular momentum, torque, and force. The octonion space $\mathbb{O}$ can be decomposed into some subspaces independent of each other, including $\mathbb{H}_g$ and $\mathbb{H}_{em}$ . The quaternion space $\mathbb{H}_g$ is able to explore the physical properties of gravitational fields. And the second subspace $\mathbb{H}_{em}$ is capable of studying the physical properties of electromagnetic fields.

In the octonion space $\mathbb{O}$ , when the octonion coordinate system is transformed rotationally, the vector part of the octonion physical quantity will vary, while the scalar part of the octonion physical quantity may remains unchanged. According to the property of rotational transformations of octonion coordinate systems, the scalar parts of octonion physical quantities are invariants. These invariants can be utilized as the basic postulates of Galilean transformation and Lorentz transformation and others.

The rotational transformations of octonion coordinate systems is capable of inferring a few invariants, relevant to the electromagnetic and gravitational fields, in a vacuum without considering the contribution of any material medium. By multiplying the basis vector $\textbf{I}_0$ by the octonion space $\mathbb{O}( r_j , R_j )$ from the left, it is able to achieve another octonion space $\mathbb{O}_u( R_j , r_j )$, according to the multiplication of octonions. a) From the octonion field equations in the octonion space $\mathbb{O}( r_j , R_j )$, one can deduce a few invariants, including the mass conservation law, energy conservation law, and power conservation law and others. b) From the octonion field equations in the octonion space $\mathbb{O}_u( R_j , r_j )$, one can infer several invariants, including the charge conservation law and others.

Strictly speaking, the mass conservation law and charge conservation law cannot be valid simultaneously. a) In the rotational transformations of the coordinate systems of octonion space $\mathbb{O}( r_j , R_j )$, the mass is conserved. The term related to the electric charge is a vector, which will vary in the rotational transformations of the octonion coordinate systems. b) In the rotational transformations of the coordinate systems of another octonion space $\mathbb{O}_u( R_j , r_j )$, the electric charge is conserved. The term related to the mass is a vector, which will change in the rotational transformations of the octonion coordinate systems. Further more, in the octonion space $\mathbb{O}( r_j , R_j )$, there are merely the invariants related to the gravitational fields. Inversely, in another octonion space $\mathbb{O}_u( R_j , r_j )$, there are only the invariants relevant to the electromagnetic fields. The two different sets of invariants are incompatible, in the Tables 1 and 2.

The octonion field strength and angular momentum can be combined together to become the octonion composite field strength, which is the octonion field strength within the material media. Further it is able to describe the physical properties of electromagnetic and gravitational fields within the material media, including the octonion field strength, field source, linear momentum, angular momentum, torque, and force within the material media.

In the rotational transformations, the scalar part of one octonion remains unchanged, so it is capable of inferring a few invariants within the material media. a) In the octonion space $\mathbb{O}( r_j , R_j )$, it is able to achieve some invariants from the octonion field equations within the material media, including the mass conservation law, energy conservation law, and power conservation law and others within the material media. b) In another octonion space $\mathbb{O}_u( R_j , r_j )$, it is able to attain several invariants from the octonion field equations within the material media, including the charge conservation law and others within the material media. Apparently, the mass conservation law and charge conservation law cannot be effective simultaneously.

The above result reveals that the invariants can be divided into two different types of groups, in the electromagnetic and gravitational fields within the material media. In the same group, multiple invariants within the material media can be established simultaneously. However, the invariants of two different types of groups within the material media are incompatible, in the Tables 3 and 4. For example, the mass conservation law and energy conservation law belong to the same group within the material media, and they are able to be effective simultaneously. But the mass conservation law and charge conservation law within the material media belong to different types of groups, so they are unable to be valid simultaneously. And the energy conservation law and charge conservation law within the material media are unable to be effective simultaneously.

Obviously, the number of groups of invariants is related to that of fundamental fields. In the electromagnetic and gravitational fields, the invariants can be divided into two different types of groups. In each of two different types of groups, multiple invariants can be effective simultaneously. The invariants of the two different types of groups cannot be valid simultaneously.

It is worth noting that although this paper only discussed some simple cases of invariants, relevant to the electromagnetic and gravitational fields in the octonion spaces, it has clearly shown that the invariants can be divided into two different types of groups, in the electromagnetic and gravitational fields. In particular, the mass conservation law and energy conservation law are able to be effective simultaneously. But the mass conservation law and charge conservation law can not be valid simultaneously. Either the energy conservation law and charge conservation law are unable to be effective simultaneously. In the future study, we shall research some influencing factors that the mass conservation law and charge conservation law can not be established simultaneously. It is planned to verify that the mass conservation law and charge conservation law are unable to be valid simultaneously, and their influence on other physical quantities.

\section*{Acknowledgements}

The author is indebted to the anonymous referees for their valuable comments on the previous manuscripts. He appreciates helpful discussions with Mr. Lei Weng on the mathematic physics. This project was supported partially by the National Natural Science Foundation of China under grant number 60677039.

\end{document}